\documentclass[aps,prl,twocolumn,groupedaddress,preprintnumbers]{revtex4}

\def\mylimit#1{\mathrel{\mathop{\kern0pt\longrightarrow}\limits_{#1}}}

\makeatletter
\@addtoreset{equation}{section}
\makeatother

\newcommand{\VEV}[1]{\left\langle #1 \right\rangle}

\newcommand{\bequ}{\begin{equation}}
\newcommand{\eequ}{\end{equation}}
\newcommand{\beqn}{\begin{eqnarray}}
\newcommand{\eeqn}{\end{eqnarray}}
\newcommand{\bctr}{\begin{center}}
\newcommand{\ectr}{\end{center}}

\begin{document}
\preprint{\vbox{\hbox{hep-ph/0402224}
\hbox{KUNS-1900}
\hbox{\today}}}

\title{$E_6$ Unification, Large Neutrino Mixings, and SUSY Flavor Problem
}

\author{Nobuhiro {\sc Maekawa}}
\email[]{maekawa@gauge.scphys.kyoto-u.ac.jp}
\affiliation{Department of Physics, Kyoto University, Kyoto 606-8502, Japan}

\date{\today}

\begin{abstract}
This study indicates that in $E_6$ grand unified theories (GUTs), 
once a hierarchical 
structure of up-quark-type Yukawa couplings is given as a basic structure of
flavor, larger lepton mixings than the quark mixings and milder 
down-quark-type 
(and charged-lepton-type) mass hierarchies than up-quark-type mass 
hierarchy can generically be obtained under a few natural assumptions. 
The basic flavor structure is compatible with non-Abelian horizontal symmetry,
which can solve the SUSY flavor problem. It is shown that in solving the 
SUSY flavor
problem, the $E_6$ structure, which
realizes bi-large neutrino mixings, also solves a problem that results from
the large neutrino mixing angles.
\end{abstract}


\maketitle


\section{Introduction}
Several recent neutrino experiments have reported that 
the neutrino mixings 
are quite large \cite{atmos,solar}, 
which are much different from those in the quark sector.
Based on the fact that these results are different from previous
ideas that the mixings in the lepton sector must be small as those
in the quark sector, various grand unified scenarios have been
examined in the literature
\cite{NGUT,Nomura,maekawa,Bando,BM,nonabel,horizontal}. 

One of the key observations in understanding
the difference between quark mixings and lepton mixings is that
quark mixings are determined by the diagonalizing matrices of 
the representation field ${\bf 10}(\ni Q=(U_L,D_L), U_R^c, E_R^c)$ 
of $SU(5)$ that includes doublet
quark $Q$, whereas lepton mixings are determined by those of 
${\bf \bar 5}(\ni D_R^c, L=(N_L, E_L))$ 
that includes doublet lepton $L$. Therefore, in the context of $SU(5)$ 
grand unified theories (GUTs), it is not difficult to obtain different 
mixing matrices of quarks and leptons \cite{NGUT}.

$SO(10)$ unification is interesting
because all one generation quarks and leptons, including the right-handed 
neutrino $N_R^c$,
can be unified into a single multiplet ${\bf 16}$, which is divided as
${\bf 16}\rightarrow {\bf 10}+{\bf \bar 5}+{\bf 1}$ under 
$SO(10)\supset SU(5)$. However, because $SO(10)$
includes $SU(2)_R$, which rotates right-handed quarks $(U_R^c, D_R^c)$ and 
right-handed leptons $(N_R^c, E_R^c)$, $SO(10)$ symmetric Yukawa interactions, 
for example, $Y_{ij}{\bf 16}_i{\bf 16}_j{\bf 10}_H$ ($i=1,2,3$) with a Higgs 
field ${\bf 10}_H$, lead to the same Yukawa 
matrices for
up and down quark sectors. However, these are neither realistic nor
consistent with non-vanishing values
of the quark mixings in the Cabibbo-Kobayashi-Maskawa (CKM) matrix. 
Therefore, to obtain
realistic mass matrices, we have to pick up a non-vanishing vacuum 
expectation value (VEV) 
of a Higgs field $C$ (representation {\bf 16} is the simplest one), 
which breaks the $SU(2)_R$, to obtain the realistic Yukawa
couplings. 
Moreover, in the high-scale supersymmetry (SUSY) breaking scenario, 
because of $SO(10)$ relations $Y_u=Y_{\nu_D}$, where $Y_{\nu_D}$ is
the neutrino Yukawa matrix for Dirac neutrino masses and $Y_u$ is
the up-quark-type Yukawa matrix one of whose components is large 
 to obtain large top-quark mass,
large mixings in lepton
sector generically result in too large $\mu\rightarrow e\gamma$ process 
through loop corrections \cite{FCNC}. 
One of the most effective ways to avoid these problems is to introduce 
several matter fields 
${\bf 10}_a$ $(a=1,\cdots,n)$ of $SO(10)$, which are divided as 
${\bf 10}\rightarrow {\bf 5}+{\bf \bar 5}$ under $SO(10)\supset SU(5)$, 
in addition to three ${\bf 16}_i$. Then the mass matrix
of $n$ ${\bf 5}$s and $n+3$ ${\bf \bar 5}$s can pick up the VEV of $C$
through the interactions ${\bf 16}_i {\bf 10}_a C$ if we take $C$ as
${\bf 16}$ of $SO(10)$ whose VEV breaks $SO(10)$ into $SU(5)$. 
This structure not only avoids the relation 
$Y_{\nu_D}=Y_u$ but also realizes
the difference between the mixing matrices of the quark and lepton sectors.
Actually, large neutrino mixings, small CKM 
mixings, and small Dirac neutrino masses can be obtained in some models
\cite{maekawa} using this mechanism.

Then, $E_6$ unification becomes more interesting, because the fundamental 
representation ${\bf 27}$ includes ${\bf 10}$ as well as ${\bf 16}$ of 
$SO(10)$ as 
\begin{equation}
{\bf27} \rightarrow \underbrace{[{\bf 10} +{\bf \bar 5}+{\bf 1}]}_{\bf 16}
+\underbrace{[{\bf \bar 5}+{\bf 5}]}_{\bf 10}
+ \underbrace{[{\bf 1}]}_{{\bf 1}}
\end{equation}
under $E_6\supset SO(10)\supset SU(5)$. Here,
the representations of $SO(10)$ and $SU(5)$ are explicitly denoted.
In the literature, it has been shown that it is possible to obtain 
not only large neutrino mixing for the atmospheric neutrino anomaly \cite{Bando}
 but also
that for the solar neutrino problem \cite{BM} in $E_6$ unification.

It is noteworthy that in Refs. \cite{maekawa}, \cite{BM}, and \cite{MY} 
generic interactions
(even for higher dimensional interactions) are introduced with $O(1)$ 
coefficients not only for the Yukawa interactions \cite{BM} but also for 
the Higgs potentials \cite{MY}.
Therefore, in the scenario, not only the Yukawa interactions but also the 
scales of VEVs are determined only by the symmetry of the models.

However, most of the above-mentioned studies concentrated on investigating
whether
it is possible to obtain large mixings in lepton sector and small mixings
in quark sector and did not investigate the reasons for which the lepton 
sector has larger mixings than the quark sector. One of the purposes of this 
paper is to clarify this phenomenon. We attribute this phenomenon
to ``$E_6$ unification." 

Moreover, we emphasize that in the $E_6$ GUTs, introducing non-Abelian 
horizontal 
symmetry $SU(2)_H$ or $SU(3)_H$ naturally solves the SUSY 
flavor problem. That is, the realistic hierarchical structure of quark and 
lepton
mass matrices can be obtained, and almost universal scalar fermion masses,
which are important in suppressing flavor changing neutral current (FCNC) 
processes, are obtained because at least the first two generation fields
are unified into a single multiplet.  If $SU(3)_H$ is adopted as the horizontal 
symmetry, 
all three generation quarks and leptons can be unified into a single multiplet
$({\bf 27},{\bf 3})$. 
Of course, the idea that non-Abelian horizontal symmetry is introduced to solve
the SUSY flavor problem is not new \cite{nonabel,horizontal,Snonabel};however, 
this paper emphasizes the fact that
the structure peculiar to $E_6$ GUTs, which realizes bi-large neutrino mixings,
 also plays an important 
role in suppressing the FCNC processes sufficiently.

The essential arguments of this paper are similar to those in our previous 
papers \cite {BM,horizontal} in which 
anomalous $U(1)$ symmetry \cite{U1} is adopted. 
However, in this paper, we show that most of the 
arguments can be generally applied to $E_6$ GUT even without anomalous
$U(1)_A$ symmetry and independent of the origin of the Yukawa hierarchy and 
the mechanism for determination of VEVs.
Further, we examine the conditions to realize the above situations.

\section{Basic assumption}
$E_6$ unification \cite{e6} is considered to be attractive 
because the gauge
anomaly is automatically free, and $E_6$ is the maximal exceptional group 
that has complex representation. Further, all the basic fermions of three 
generations 
are unified into three fundamental
representation fields $\Psi_i({\bf 27})$, which are divided as
\begin{equation}
\Psi_i({\bf27}) \rightarrow {\bf 16}_{\Psi_i}+{\bf 10}_{\Psi_i}
+{\bf 1}_{\Psi_i}
\end{equation}
under $E_6\supset SO(10)$.

It has been argued that the additional component fields ${\bf 5}$ and 
${\bf \bar 5}$ of $SU(5)$ play an important role in realizing rich structure 
in Yukawa couplings \cite{Nano} and in realizing
large neutrino mixings \cite{Bando,BM,stech}, 
because three of six
${\bf \bar 5}$ fields become superheavy and light fields are linear 
combinations of these six ${\bf \bar 5}$ fields. In order to estimate the 
$3\times 6$ mass matrix of three ${\bf 5}$ fields and six ${\bf \bar 5}$ 
fields, we have to fix a part of the Higgs sector that breaks $E_6$ 
into the standard
model (SM) gauge group
$G_{\rm SM}=SU(3)_C\times SU(2)_L\times U(1)_Y$. Supposing that Higgs fields,
$\Phi({\bf 27})$ and $\bar \Phi({\bf\overline{27}})$, break $E_6$ into $SO(10)$ 
($|{\VEV{{\bf 1}_\Phi}}|=|\VEV{{\bf 1}_{\bar\Phi}}|$ to satisfy the $D$-flatness
conditions),
$C({\bf 27})$ and $\bar C({\bf\overline{27}})$ break $SO(10)$ into $SU(5)$ 
($|{\VEV{{\bf 16}_C}}|=|\VEV{{\bf\overline{16}}_{\bar C}}|$ to satisfy
the $D$-flatness conditions),
and an adjoint Higgs $A({\bf 78})$ breaks $SU(5)$ into 
the standard model (SM) gauge group 
$G_{\rm SM}=SU(3)_C\times SU(2)_L\times U(1)_Y$. (Here, we do not fix the 
direction of the
VEV of $\VEV{A}$ and the scale of 
$\VEV{A}$, which may be larger than the VEVs of $\Phi$ or $C$, 
because it is independent of
the following arguments.) Then, through the interactions
\begin{equation}
W_Y=Y^\Phi_{ij}\Psi_i\Psi_j\Phi+Y^C_{ij}\Psi_i\Psi_jC,
\label{yukawa}
\end{equation}
the mass matrix of ${\bf 5}$ and ${\bf \bar 5}$ is determined by 
developing the VEVs of $\Phi$ and $C$.

In the following calculations, for simplicity, we assume that the main 
modes of the SM 
doublet Higgs $H_u$ and $H_d$ come from ${\bf 10}_\Phi$ and ${\bf 10}_C$.
 Therefore, the Yukawa couplings for the up-quark
sector are essentially determined by the Yukawa couplings $Y^\Phi$ and $Y^C$.
However, including the Higgs mixings with ${\bf 16}_\Phi$ or
${\bf 16}_C$ does not change the following conclusion drastically unless these
${\bf 16}$ components dominate the ${\bf 10}$ components.

One of the most important basic assumptions is as follows.
The Yukawa matrices $Y^\Phi$ and $Y^C$ have a hierarchical structure
that can realize the hierarchy in the up-quark sector.  
In the literature \cite{FN,nonabel,extra,Inoue,Nelson}, several 
mechanisms have been proposed to understand the Yukawa hierarchies. 
Here, we do not fix the mechanism that realizes such a hierarchical structure.
However, we simply assume that both the hierarchies of each of the two 
Yukawa matrices,
$Y^\Phi$ and  $Y^C$, have the same origin,
that is,
\begin{equation}
{(Y^C)_{ij}}\sim {(Y^\Phi)_{ij}}\equiv Y_{ij}.
\end{equation}
More precisely, we assume that the order of each component 
of $Y^\Phi$ is the same as 
that of the corresponding component of $Y^C$, but generally 
$(Y^\Phi)_{ij}\neq (Y^C)_{ij}$.

\section{Quark and Lepton Mass Matrices}
For simplicity, in the following argument, we adopt
\begin{equation}
Y_{ij}\sim \left(\matrix{0 & \lambda^5 & 0 \cr
                         \lambda^5 & \lambda^4 & \lambda^2 \cr
                         0 & \lambda^2 & 1 \cr}\right){\rm or}
           \left[\left(\matrix{\lambda^6 & \lambda^5 & \lambda^3 \cr
                         \lambda^5 & \lambda^4 & \lambda^2 \cr
                         \lambda^3 & \lambda^2 & 1 \cr}\right)\right],
\end{equation}
where we take $\lambda\sim \sin \theta_C \sim 0.22$. The former
is called type A and the latter is type B in this paper.
Note that the following arguments can be applied to various other types of 
matrices that can obtain a realistic up-type quark mass matrix.
Once we fix this basic structure of the Yukawa couplings $Y$, 
the mass matrix of 
$({\bf 5, \bar 5})$ fields is given by
\begin{eqnarray}
M_{({\bf 5,\bar 5})}&=&(Y^\Phi\VEV{{\bf 1}_\Phi}, Y^C\VEV{{\bf 16}_C})
=\VEV{{\bf 1}_\Phi}(Y^\Phi,RY^C),\nonumber \\
R&\equiv& \VEV{{\bf 16}_C}/\VEV{{\bf 1}_\Phi}\equiv \lambda^r.
\end{eqnarray}
As discussed in Ref. \cite{BM}, 
the light ${\bf\bar 5}$ fields can be
classified by the parameter $r$. If the VEV $\VEV{\Phi}$ is not much larger
than the VEV $\VEV{C}$ ($0\leq r < 3$), then all three main modes of the 
light ${\bf \bar 5}$ fields come from  
the first two generation fields, $\Psi_1$ and $\Psi_2$.
That is, the main light modes become 
${\bf 16}_{\Psi_1}$, ${\bf 10}_{\Psi_1}$, and 
${\bf 16}_{\Psi_2}$.
Using the basis in which each light mode includes no other
main light modes, the light modes can be written as
\begin{eqnarray}
{\bf \overline 5}_1 &=& {\bf 16}_{\Psi_1}
+\lambda^3{\bf 16}_{\Psi_3}
[+\lambda^{2+r}{\bf 10}_{\Psi_2}]
+\lambda^{3+r}{\bf 10}_{\Psi_3}, 
\nonumber \\
{\bf \overline 5}_2 &=& {\bf 10}_{\Psi_1}
+\lambda^{3-r}{\bf 16}_{\Psi_3}
[+\lambda^2{\bf 10}_{\Psi_2}]
+\lambda^3{\bf 10}_{\Psi_3}, 
\label{mix} \\
{\bf \overline 5}_3 &=& {\bf 16}_{\Psi_2}
+\lambda^2{\bf 16}_{\Psi_3}
+\lambda^r{\bf 10}_{\Psi_2}
+\lambda^{2+r}{\bf 10}_{\Psi_3},
\nonumber
\end{eqnarray}
where the first terms on the right-hand sides are the main components of
these massless modes, and the other terms are mixing terms with
the heavy states, ${\bf 16}_{\Psi_3}$,
${\bf 10}_{\Psi_2}$, and
${\bf 10}_{\Psi_3}$. Note that the coefficients of mixings for all three heavy 
modes 
are essentially determined by the hierarchical 
Yukawa couplings $Y$ and $R$.
Further, the ${\bf 10}_{\Psi_i}$ modes have no Yukawa couplings 
if the SM doublet
Higgs comes from ${\bf 10}_\Phi$ and ${\bf 10}_C$. Therefore, the mixings
of ${\bf 16}_{\Psi_3}$ are important to estimate the Yukawa couplings.
Then, Yukawa couplings can be calculated as
\begin{eqnarray}
Y_u&\sim & Y, \nonumber \\
Y_d&\sim & Y_e^T \\
&\sim&
\left(\matrix{Y_{11}+Y_{13}\lambda^3 
                  & Y_{13}\lambda^{3-r} 
                  & Y_{12}+Y_{13}\lambda^2 \cr
                   Y_{21}+Y_{23}\lambda^3 
                   & Y_{23}\lambda^{3-r} 
                   & Y_{22}+Y_{23}\lambda^2 \cr
                   Y_{31}+Y_{33}\lambda^3 & 
                   Y_{33}\lambda^{3-r} 
                   & Y_{32}+Y_{33}\lambda^2 }
                   \right). \nonumber
\end{eqnarray}
It is apparent that the Yukawa matrix $Y_d$ has a milder hierarchy than
that of $Y_u$ under these conditions. 
The essential point is that the ${\bf \bar 5}$
fields from $\Psi_3$, which has larger Yukawa couplings, become superheavy,
and the light ${\bf \bar 5}$ fields come from the first two generation 
fields $\Psi_1$ and $\Psi_2$.
This results in a small 
$\tan \beta\equiv \VEV{H_u}/\VEV{H_d}$, which is roughly estimated as
$\tan \beta \sim (m_t/m_b)(Y_{32}/Y_{33})$ up to renormalization
group effects, because it is expected that $(Y_{32}/Y_{33})^2\leq m_c/m_t$.

If we take $r=0.5$, 
the Yukawa matrices become
\begin{equation}
Y_d\sim Y_e^T
\sim \left(\matrix{0 [\lambda^6] & 0 [\lambda^{5.5}] & \lambda^5 \cr
                   \lambda^5 & \lambda^{4.5} & \lambda^4 \cr
                   \lambda^3 & \lambda^{2.5} & \lambda^2 }
                   \right), 
\end{equation}
which give almost realistic masses of down quarks and charged leptons
when 
$\tan\beta\sim (m_t/m_b)\lambda^2$ up to renormalization effects.
The CKM matrix can be calculated as
\begin{equation}
V_{CKM}\sim \left(\matrix{1 & \lambda & \lambda^3 \cr
                          \lambda & 1 & \lambda^2 \cr
                          \lambda^3 & \lambda^2 & 1 \cr
                   }\right).
\end{equation}

In order to avoid the unrealistic $SU(5)$ GUT relation, $Y_d=Y_e^T$,
we have to pick up the VEVs of $A$ in these Yukawa matrices. In principle, 
we can pick up the effects in the mass matrix of ${\bf 5}$ and ${\bf \bar 5}$
and/or Yukawa interactions by the higher dimensional interactions. Here, we 
simply assume it in order to obtain realistic quark and lepton mass matrices. 
This point is discussed in the following sections.

\section{Neutrino masses and mixings}
Because representation ${\bf 27}$ has two singlets $N_R^c$ and $S$ 
under the SM gauge group,
the Dirac mass matrix becomes $3\times 6$. 
The Dirac neutrino mass matrix is obtained from the interactions (\ref{yukawa})
as
\begin{equation}
Y_{\nu_D}=(Y_N,Y_S)\sim (1,\lambda^r)\otimes \left(\matrix{
               0[\lambda^6] & \lambda^5 & \lambda^3 \cr
               0[\lambda^{6-r}] & \lambda^{5-r} & \lambda^{3-r} \cr
               \lambda^5 & \lambda^4 & \lambda^2\cr}\right) 
\end{equation}
Because the Dirac neutrino masses become $\lambda^2$ smaller than the usual 
$SO(10)$ GUT predictions $Y_u\sim Y_{\nu_D}$, 
the smaller right-handed neutrino masses are required
to realize the correct neutrino mass scales corresponding to the observed 
neutrino oscillations. (Note that this difference is 
important in suppressing FCNC processes (e.g. $\mu\rightarrow e\gamma$)
that originate from loop corrections to SUSY breaking parameters \cite{FCNC}, 
because such
corrections are proportional to $Y_{\nu_D}^\dagger Y_{\nu_D}$.)
The right-handed neutrino mass matrix ($6\times 6$) is obtained from the 
interactions
\begin{equation}
Y^{\bar X\bar Y}_{ij}\Psi_i\Psi_j\frac{\bar X\bar Y}{\Lambda},
\end{equation}
where $\bar X, \bar Y=\bar \Phi, \bar C$,
as
\begin{eqnarray}
M_{\nu_R}&=&\left(\matrix{Y^{\bar\Phi\bar\Phi}\VEV{\bar\Phi}^2 & 
                        Y^{\bar\Phi\bar C}\VEV{\bar\Phi}\VEV{\bar C} \cr
                        Y^{\bar\Phi\bar C}\VEV{\bar\Phi}\VEV{\bar C} &
                        Y^{\bar C\bar C}\VEV{\bar C}^2 }\right)
                        \frac{1}{\Lambda}, \nonumber \\
         &\sim& \left(\matrix{ 1 & \lambda^r \cr
                              \lambda^r & \lambda^{2r} \cr}\right)\otimes
         \left(\matrix{0[\lambda^6] & \lambda^5 & 0[\lambda^3] \cr
                       \lambda^5 & \lambda^4 & \lambda^2 \cr
                       0[\lambda^3] & \lambda^2 & 1 \cr}\right)
                       \frac{c\VEV{\bar \Phi}^2}{\Lambda}.
\end{eqnarray}
Here, we take
$Y^{\bar X\bar Y}\sim c Y^\Phi$,
where $c$ is a constant.
Note that the smallness of the right-handed neutrino masses is naturally 
expected
in this scenario, because the right-handed neutrino masses are obtained from 
the higher dimensional interactions.
Then, the light neutrino mass matrix is obtained by seesaw mechanism 
\cite{seesaw} as
\begin{eqnarray}
M_{\nu}&=&Y_{\nu_D}M_{\nu_R}^{-1}Y_{\nu_D}^T\VEV{H_u}^2\eta^2 \nonumber \\
&\sim &
\lambda^4\left(\matrix{\lambda^2 & \lambda^{2-r} & \lambda \cr
                       \lambda^{2-r} & \lambda^{2-2r} & \lambda^{1-r} \cr
                       \lambda & \lambda^{1-r} & 1\cr}\right)
        \frac{\Lambda\VEV{H_u}^2\eta^2}{c\VEV{\bar\Phi}^2},
\end{eqnarray}
where $\eta$ is a renormalization parameter. 
If we take $\VEV{\Phi}\sim 10^{16}$ GeV, $\Lambda\sim 10^{19}$ GeV, 
$\VEV{H_u}\eta\sim 100$ GeV,
$c\sim 0.1$, and $r\sim0.5$, we can obtain realistic neutrino masses and 
the Maki-Nakagawa-Sakata (MNS) matrix \cite{MNS} as
\begin{equation}
V_{MNS}\sim \left(\matrix{ 1 & \lambda^{0.5} & \lambda \cr
                          \lambda^{0.5} & 1 & \lambda^{0.5} \cr
                          \lambda & \lambda^{0.5} & 1 \cr}\right).
\end{equation}

\section{Non-Abelian horizontal symmetry}
One of the most attractive features in this $E_6$ GUT is that non-Abelian
horizontal symmetry can be naturally introduced to solve the SUSY flavor
problem. One of the important points in the above scenario is that all 
the realistic mass
hierarchies of quarks and leptons can be obtained from just one basic 
hierarchical 
structure of the Yukawa matrix, $Y$, which can be naturally obtained from one 
non-Abelian horizontal symmetry. (Usually, to obtain various hierarchies of
quarks and leptons, several non-Abelian horizontal symmetries are introduced
or the coefficients are tuned to realize various hierarchies from one
horizontal symmetry, or more Higgs fields whose VEVs break the horizontal 
symmetry are introduced.)

Usually, $SU(2)_H$ or $U(2)_H$ horizontal symmetry is adopted because
it realizes the universal sfermion masses of the first two generation 
fields that are doublets under 
the horizontal symmetry, which is important in solving the SUSY flavor 
problem, 
and because top Yukawa coupling is allowed by the symmetry when the 
third generation
fields and the Higgs fields are singlets under the horizontal symmetry.

However, the universal sfermion masses of the first two generation fields,
$\tilde m_{\bf \bar 5}^2\sim 
{\rm diag}(\tilde m^2, \tilde m^2, (a+1)\tilde m^2)$, are not enough 
to suppress the FCNC processes in the GUT models in which 
diagonalizing matrices $V_x$ $x={\bf \bar 5},{\bf 10}$ 
can be estimated as
$V_{\bf 10}\sim V_{CKM}$ and $V_{\bf \bar 5}\sim V_{MNS}$
as in the previous models. 
This is because 
the mixing matrices defined as 
$\delta_{x}\equiv V_{x}^\dagger 
\frac{\tilde m_{x}^2-\tilde m^2}{\tilde m^2}V_{x}$\cite{GGMS} become
\begin{equation}
\delta_{\bf\bar 5}\sim V_{\bf \bar 5}^\dagger\left(
  \matrix{ 0 & 0 & 0 \cr
           0 & 0 & 0 \cr
           0 & 0 & a \cr}\right)V_{\bar 5}\sim
   \left(\matrix{\lambda^2 & \lambda^{1.5} & \lambda \cr
           \lambda^{1.5} & \lambda & \lambda^{0.5}\cr
           \lambda & \lambda^{0.5} & 1\cr}\right)a
\end{equation}
at the GUT scale, which does not satisfy 
the constraints of various FCNC processes;
\begin{eqnarray}
\sqrt{|{\rm Im}(\delta_{D_L})_{12}(\delta_{D_R})_{12})|}&\leq&
 2\times 10^{-4}\left(\frac{\tilde m_Q}{500\ {\rm GeV}}\right) \nonumber \\
|{\rm Im}(\delta_{D_R})_{12}| & \leq & 1.5\times 10^{-3}
\left(\frac{\tilde m_Q}{500\ {\rm GeV}}\right),
\label{K}
\end{eqnarray}
at the weak scale from $\epsilon_K$ in $K$ meson mixing,
and 
\begin{equation}
|(\delta_{E_L})_{12}|\leq 4\times 10^{-3}\left(
\frac{\tilde m_L}{100\ {\rm GeV}}\right)^2
\label{mu}
\end{equation}
from the $\mu\rightarrow e\gamma$ process.

It is interesting that in the $E_6$ unification, such a problem is 
naturally solved. As discussed in the previous section, in $E_6$ GUT, 
it is natural that all the three light ${\bf \bar 5}$ fields come from the 
first two generation fields. Therefore, if we introduce horizontal symmetry
$SU(2)_H$ or $U(2)_H$, all the sfermion masses of ${\bf \bar 5}$ fields
become equivalent in the leading order, that is, $a=0$. Note that the structure
also realizes bi-large neutrino mixings as discussed in the previous section.
It is suggestive that the same structure solves the FCNC problem that originate
from the large neutrino mixings.

Of course, the horizontal symmetry has to be broken to obtain the realistic
quark and lepton mass matrices, and this effect also breaks the universality
of the sfermion masses.  
Next, we examine this breaking effect 
in a concrete model. 

We introduce a global horizontal symmetry $U(2)_H$ and 
fields listed in Table I.
\begin{center}
Table I. Odd R-parity for matter fields $\Psi$ and $\Psi_3$
is introduced.

\begin{tabular}{|c|c|c|c|c|c|c|c|c|c|c|} 
\hline
   &$\Psi$&$\Psi_3$&$\bar F$&$\Theta$&    $A$ & 
   $\Phi$ & $\bar \Phi$ & $C$ & $\bar C$  \\
\hline 
 $E_6$  &{\bf 27}&{\bf 27}&{\bf 1}&{\bf 1}&{\bf 78}  
 & {\bf 27} & ${\bf \overline{27}}$ &{\bf 27} & ${\bf \overline{27}}$ \\
 $SU(2)_H$ & {\bf 2} & {\bf 1}&${\bf \bar 2}$& {\bf 1}& {\bf 1} 
 &{\bf 1}& 
 {\bf 1} & {\bf 1} & {\bf 1}   \\
 $U(1)_H$  & 1 & 0 & -1 & -2 & 0 & 0 & 0 & 0 & 0  \\
 \hline
\end{tabular}
\end{center}
Here, we simply assume that $U(2)_H$ is broken by the VEVs of $\bar F$ and 
$\Theta$ as
\begin{equation}
  U(2)_H \mylimit{\VEV{\Theta}\sim \lambda}SU(2)_H 
  \mylimit{\VEV{\bar F_a}\sim \lambda^2\delta_a^2} 
  {\rm nothing},
\end{equation}
and $E_6$ is broken into the SM gauge group by the VEVs
of $\Phi$, $\bar \Phi$, $C$, $\bar C$, and $A$;
\begin{eqnarray}
\VEV{{\bf 1}_\Phi}&=&\VEV{{\bf 1}_{\bar \Phi}}\sim \lambda^{3.5}, \\
\VEV{{\bf 16}_C}&=&\VEV{{\bf \overline{16}}_{\bar C}}\sim \lambda^4,\\
\VEV{A}&\sim& \lambda^4.
\end{eqnarray}
In this paper, we sometimes use a unit in which $\Lambda=1$. 
Then, the type A basic hierarchical Yukawa matrix $Y$ 
is obtained from 
the interactions
\begin{equation}
((\Psi_3)^2+\Psi_3\Psi \bar F+(\Psi\bar F)^2+\Psi A\Psi\Theta)(\Phi+C).
\label{interaction}
\end{equation}
Note that the terms $\Psi\Psi\Theta(\Phi+C)$ become trivially zero.
The light ${\bf\bar 5}$ fields are obtained as in Eq. (\ref{mix}) with
$r=0.5$. Here, for simplicity, $H_d$ comes from a linear combination 
${\bf 10}_\Phi+{\bf 10}_C$. (If both SM Higgs $H_d$ and $H_u$ come from 
only ${\bf 10}_\Phi$, CKM mixing becomes too small
because of a cancellation.) Then, we obtain a realistic quark and lepton 
mass matrices, including bi-large neutrino mixings,  as discussed in the 
previous section.
Here, the interaction $\Psi A\Psi\Theta(\Phi+C)$ plays an important role 
in avoiding
unrealistic $SU(5)$ GUT relations $Y_d=Y_e^T$. Generically, the VEV of $A$
gives different contributions to the Yukawa couplings of quarks and leptons;
therefore, the quark mass matrices can be different
from the lepton mass matrices. Note that the mixing 
coefficients of $D_R^c$ in 
Eq. (\ref{mix}) are of the same order as those of $L$ but have generically 
different values from those of $L$.

Note that the main modes of the light ${\bf \bar 5}$ fields
are obtained from the first two generation fields, which come from 
a single field $\Psi$ in this case. Therefore, unless the horizontal 
symmetry is broken,
the sfermion masses are obtained as
\begin{eqnarray}
\tilde m_{\bf 10}^2&\sim &\tilde m^2\left(\matrix{1 & 0 & 0 \cr
                                                0 & 1 & 0 \cr
                                                0 & 0 & O(1) \cr}\right),\\
\tilde m_{\bf\bar 5}^2&\sim  &\tilde m^2
\left(\matrix{1+\lambda^6 & \lambda^{5.5} & \lambda^5 \cr
              \lambda^{5.5} & 1+\lambda^5 & \lambda^{4.5} \cr
              \lambda^5 & \lambda^{4.5} & 1+\lambda^4 \cr}\right),
\end{eqnarray}
where the corrections to the sfermion masses $\tilde m_{\bf \bar 5}$ come
from the mixings in Eq. (\ref{mix}). 
When the breaking of the horizontal symmetry is taken into account,
the sfermion mass matrices are corrected as
\begin{eqnarray}
\Delta\tilde m_{\bf 10}^2
&\sim &\tilde 
m^2\left(\matrix{\lambda^4 & \lambda^5 & \lambda^3 \cr
                 \lambda^5 & \lambda^4 & \lambda^2 \cr
                 \lambda^3 & \lambda^2 & O(1) \cr}\right), \\
\Delta\tilde m_{\bf\bar 5}^2
&\sim &\tilde m^2\left(\matrix{\lambda^4 & \lambda^{6.5} & \lambda^5 \cr
                             \lambda^{6.5} & \lambda^4 & \lambda^{4.5} \cr
                             \lambda^5 & \lambda^{4.5} & \lambda^4 \cr}\right),
\end{eqnarray}
which are calculated mainly from the interactions
\begin{eqnarray}
\int d^2\theta d^2\bar\theta &(&(\Psi_3+\Psi \bar F+\Psi \bar F^\dagger\Theta
)^\dagger (\Psi_3+\Psi \bar F+\Psi \bar F^\dagger\Theta) \nonumber \\
&+&\Psi^\dagger A\Psi+\Psi_3^\dagger A\Psi_3)
\frac{X^\dagger X}{\Lambda^2}
\label{SUSY}
\end{eqnarray}
where $X$ is a spurion field whose VEV of $F$-term is given as 
$F_X=\tilde m\Lambda$.
The last two terms in Eq. (\ref{SUSY}) splits the masses of scalar down quarks
included in fields ${\bf 10}$ and ${\bf 16}$ because they have different
$B-L$ charges. Such effects is important in estimating the corrections to 
sfermion masses because only the main mode of 
${\bf \bar 5}_2$ comes from ${\bf 10}$ of $SO(10)$ and the other main modes come
from ${\bf 16}$.

The mixing matrices 
$\delta_x\equiv V_x^\dagger \Delta \tilde m_x^2 V_x/\tilde m^2$ 
$(x={\bf \bar 5}, {\bf 10})$ are approximated as  
\begin{equation}
\delta_{\bf 10}=\left(\matrix{\lambda^4 & \lambda^5 & \lambda^{3} \cr
                          \lambda^5 & \lambda^4 & \lambda^{2} \cr
                          \lambda^{3} & \lambda^{2} & O(1) }
             \right), \ 
\delta_{\bf \bar 5}=
          \left(\matrix{\lambda^4 & \lambda^{4.5} & \lambda^{5} \cr
                          \lambda^{4.5} & \lambda^4 & \lambda^{4.5} \cr
                          \lambda^{5} & \lambda^{4.5} & \lambda^4 }
             \right)
\label{delta}
\end{equation}
at the GUT scale. The constraints (\ref{K}) at the weak scale from 
$\epsilon_K$ in $K$ meson mixing
require scalar quark masses larger than
300 GeV, because in this model
$\sqrt{|(\delta_{D_L})_{12}(\delta_{D_R})_{12})|}\sim \lambda^{4.75}
(\eta_q)^{-1}$
and $|(\delta_{D_R})_{12}|\sim \lambda^{4.5}(\eta_q)^{-1}$, where
we take a renormalization factor $\eta_q\sim 6$
\footnote{
The renormalization factor is strongly dependent on the ratio of the gaugino 
mass to the scalar fermion mass and the model below the GUT scale.
If the model is MSSM and the ratio at the GUT scale is 1, 
then $\eta_q=6\sim 7$.}.
Further, the constraint from the $\mu\rightarrow e\gamma$ process in
Eq. (\ref{mu}) 
is easily satisfied, 
because 
$|(\delta_{E_L})_{12}|\sim \lambda^{4.5}$ in this model.

It is quite impressive that the structure which realizes bi-large neutrino 
mixings simultaneously solves the SUSY FCNC problem in non-Abelian horizontal
symmetry that originates from large neutrino mixings.

If the spurion field $X$ has interactions in the superpotential
\begin{equation}
((\Psi_3)^2+\Psi_3\Psi \bar F+(\Psi\bar F)^2+\Psi A\Psi\Theta)(\Phi+C)X,
\label{A}
\end{equation}
the left-right mixings in sfermion masses $\Delta_X^{LR}$ 
$(X=U,D,L)$ are 
induced. In the above models,
$\delta_X^{LR}\equiv \Delta_X^{LR}/\tilde m_X^2$ are calculated as
\begin{eqnarray}
\delta_D^{LR}&\sim &\delta_L^{RL}
\sim \left(\matrix{\lambda^6 & \lambda^{5.5} & \lambda^5 \cr
                   \lambda^5 & \lambda^{4.5} & \lambda^4 \cr
                   \lambda^3 & \lambda^{2.5} & \lambda^2 }
                   \right)\frac{\VEV{H_d}}{\tilde m_X}, \\
\delta_U^{LR}
&\sim &\left(\matrix{\lambda^6 & \lambda^{5} & \lambda^3 \cr
                   \lambda^5 & \lambda^{4} & \lambda^2 \cr
                   \lambda^3 & \lambda^{2} & 1 }
                   \right)\frac{\VEV{H_u}}{\tilde m_U}. 
\end{eqnarray}
The constraints for these mixings
\begin{eqnarray}
|{\rm Im} (\delta_D^{LR})_{12}|&<&2\times 10^{-5}\times 
                          \left(\frac{\tilde m_d}{500\ {\rm GeV}}\right)^2,
                          \\
|(\delta_D^{LR})_{23}|&<&1.6\times 10^{-2}\times 
                          \left(\frac{\tilde m_d}{500\ {\rm GeV}}\right)^2,
                          \\
|(\delta_L^{LR})_{12}|&<&1\times 10^{-6}\times 
                          \left(\frac{\tilde m_l}{100\ {\rm GeV}}\right)^2,
                          \\
|(\delta_L^{LR})_{23}|&<&6\times 10^{-3}\times 
                          \left(\frac{\tilde m_l}{100\ {\rm GeV}}\right)^2,
\end{eqnarray}
which are obtained from $\epsilon'/\epsilon$, $b\rightarrow s\gamma$,
$\mu\rightarrow e\gamma$, and $\tau\rightarrow \mu\gamma$ processes, 
respectively \cite{GGMS}, require roughly
$\tilde m_d> 1$ TeV and $\tilde m_l> 500$ GeV. 
These constraints are not so severe especially in our scenario,
because the sfermion masses can be large except those of the third
generation ${\bf 10}_3$ of $SU(5)$ that must be small to stabilize the
weak scale. Moreover, a reasonable assumption like SUSY breaking 
in the hidden sector, can lead to vanishing $\delta^{LR}$ \cite{SW}.
Actually, if the SUSY breaking sector is separated from the visible 
sector in the superpotential, the interactions ($\ref{A}$) are forbidden.

Because there is a strong suspicion that global symmetries are broken through
quantum gravitational effects, it may be more important to use local symmetries
instead of the global symmetries. For example, we can regard the 
$SU(2)_H\times U(1)_H$ symmetry in Table I as local symmetry, if we add a field 
$F({\bf 2},{\bf 1})$ whose VEV is given as $|\VEV{F}|=|\VEV{\bar F}|$ to satisfy
the $D$-flatness condition of $SU(2)_H$. (Here, the $D$-flatness
condition of $U(1)_H$ seems not to be satisfied; however, 
the Fayet-Illiopoulos $D$-term, 
if any, can improve the situation. Further, $U(1)_H$ may be 
anomalous $U(1)$, whose anomaly is cancelled by Green-Schwarz mechanism
\cite{GS}.)
This modification changes the basic structure of Yukawa couplings into 
type B, but the mixing matrices $\delta_x$ has no essential difference.
However, generally, the $D$-term of $SU(2)_H$ has non-vanishing VEV, which 
may be another source to break the universality of the sfermion masses. 
Therefore, we must assume that
the $D$ of $SU(2)_H$ is sufficiently small due to some mechanism. For example,
in the superstring theory,
if modular weights of $F$ and $\bar F$ are equivalent, the SUSY breaking
masses for the scalar components of $F$ and $\bar F$ become equal, which
realizes the vanishing $D$ of $SU(2)_H$ \cite{KMY}.  

It is straightforward to extend the horizontal symmetry $SU(2)_H$ to 
$SU(3)_H$ in which all the three family quark and leptons can be unified
into a single multiplet $({\bf 27},{\bf 3})$ under $E_6\times SU(3)_H$
as discussed in Ref. \cite{horizontal}. 
Because $SU(3)_H$ must be broken around the cutoff
scale to obtain large top Yukawa coupling, mass matrices of sfermion have
no essential difference from those of $SU(2)_H$ models.

In either of the horizontal symmetries, this scenario predicts a special 
pattern of
sfermion masses. That is, all the sfermion masses are universal except those
of the third generation ${\bf 10}_3=(Q, U_R^c, E_R^c)$ fields 
around the GUT
scale. This prediction can be tested by a linear collider in the future.

\section{Discussion}
In this paper, we did not discuss methods to determine the VEVs of the Higgs 
which break $E_6\times U(2)_H$ into the SM gauge group.
Therefore, we have simply determined the scales of non-vanishing VEVs 
of fields $F$, $\bar F$, $A$, $\Phi$, $\bar \Phi$, $C$, and $\bar C$, 
and the SM Higgs mixings
in order that realistic (scalar) fermion masses and mixings are obtained.
The SM Higgs mixing is important to avoid a cancellation of CKM mixings.
The VEV of $\bar F$ is determined in order to satisfy 
$\VEV{\bar F}\sim \sqrt{m_c/m_t}\sim \lambda^{2}$ at the GUT scale. 
It is interesting that 
this value is enough to satisfy the various FCNC constraints.
The ratio $R\equiv\VEV{C}/\VEV{\Phi}\sim \lambda^{0.5}$ is important to
obtain bi-large neutrino mixings, although $R\sim\lambda$ also gives a realistic
pattern of quark and lepton mass matrices. 
Because the $E_6$ structure is important in solving the SUSY flavor problem,
the VEVs $\VEV{\Phi}$, $\VEV{C}$, or $A$, which break $E_6$, cannot be taken 
as very large values in order to suppress FCNC processes.

If $\VEV{{\bf 45}_A}\sim \lambda^4
\left(\matrix{0& 1\cr -1 & 0}\right)\otimes
{\rm diag} (1,1,1,0,0)\propto Q_{B-L}$, which sometimes plays an important
role in solving doublet-triplet splitting problem,  is adopted, 
the mixings of lepton doublet $L$ in Eq. (\ref{mix}) are drastically
changed in the type A model used in this paper, 
because the $L$ and $\bar L$ fields in ${\bf 10}$ of $SO(10)$ have
vanishing $Q_{B-L}$ charges. In this case, other contributions to the basic
Yukawa coupling $Y$ are required. (For example, the type B model, in which
the horizontal symmetry is localized, does not change
the basic arguments in this paper because of other contributions.) 

It is obvious that our arguments can be applied to any model in which
the appropriate scales of VEVs are obtained; in other words, the appropriate
coefficients in the Higgs potential are obtained.
Therefore,  it is interesting to examine which among the various mechanisms 
that can determine coefficients of interactions are suitable
for our arguments. Such a project is important but is beyond the scope of 
this paper.

This is a promising way \cite{maekawa,BM,MY,Unif} to introduce an 
anomalous $U(1)_A$ symmetry (or
just $U(1)$ with Fayet-Illiopoulos $D$-term) to control the Higgs sector.
This is because the scales of the various non-vanishing VEVs and the SM
Higgs mixings are determined by their $U(1)_A$ charges, for example, 
$\VEV{A}\sim \lambda^{-a}$ (Here, we use small letter for their charges). 
Such determination of VEVs plays an important
role in defining the effective charges that determine all the orders of 
coefficients. Further,  because the interactions including $A$ (for example, 
$\lambda^{\psi_i+\psi_j+a+\phi}\Psi_iA\Psi_j\Phi$)  
automatically gives the same contributions to the Yukawa couplings 
as the interaction without $A$ 
($\lambda^{\psi_i+\psi_j+\phi}\Psi_i\Psi_j\Phi$),  the unrealistic GUT 
relations between fermion mass matrices can be naturally avoided. 
Moreover, such VEV relations guarantee the natural gauge coupling unification
\cite{Unif},
though it requires a rather small cutoff scale such as 
$\Lambda\sim 2\times 10^{16}$ GeV.
Actually, using anomalous $U(1)_A$ symmetry, we can obtain a complete GUT
with $E_6\times SU(3)_H$ or $E_6\times SU(2)_H$ gauge 
symmetry \cite{horizontal,MY}, in which
doublet-triplet splitting, realistic quark and lepton mass 
matrices, natural gauge coupling unification, and suppressed FCNC are realized.
Although the suppression of FCNC becomes milder than the concrete model 
discussed in this paper 
because of the lower cutoff,
it is interesting to note that such models can be built with generic 
interactions
(including higher dimensional interactions) with $O(1)$ coefficients.

$E_6$ symmetry is sufficient, but not necessary to realize the interesting 
structure discussed in this paper.
$SU(2)_E$ \cite{Bando} is the essential symmetry,  
which rotates two ${\bf\bar 5}$s and
two ${\bf 1}$s  of $SU(5)$ in a ${\bf 27}$ of $E_6$ as a doublet. 
Therefore, if a gauge group that includes $SU(2)_E$, (for example,
$SU(3)^3$ \cite{su3}, $SU(6)\times SU(2)_E$, and flipped $SO(10)$ 
[$SO(10)'\times U(1)$] \cite{flipped}) is adopted, then
the arguments in this paper can be applied.

\section{Summary}
In this paper, we pointed out that in $E_6$ GUT, once a basic 
hierarchical structure of Yukawa couplings is given, larger neutrino mixings
than the CKM mixings are naturally realized. This is
independent of the origin of the hierarchy. Therefore, this mechanism can be
applied to various models in which the hierarchical structure of Yukawa couplings
is realized. 

Moreover, we pointed out that non-Abelian horizontal symmetry is a promising
candidate for the origin of the hierarchy, which can solve the SUSY flavor 
problem. This is 
because only one hierarchical structure, which is easily obtained by introducing
a non-Abelian horizontal symmetry,  induces all the other hierarchical
structures of quark and lepton mass matrices.
It is non-trivial that in $E_6$ GUT, the structure, which realizes bi-large 
neutrino mixings, also realizes universal sfermion masses for all three 
generation ${\bf \bar 5}$ fields that are important in suppressing FCNC 
processes naturally in GUT theory with bi-large neutrino mixings.

It is quite impressive that unification of quark and leptons realizes
larger neutrino mixings and solves the SUSY flavor problem in $E_6$ 
unification. 
We hope that this compatibility is an evidence of $E_6$ GUT or the $E_6$ 
structure
in our world and the characteristic pattern of sfermion masses is confirmed
in future experiments.

\section{Acknowledgement}
N.M. thanks Z. Berezhiani and T. Yamashita for valuable discussions.
He is supported in part by Grants-in-Aid for Scientific 
Research from the Ministry of Education, Culture, Sports, Science 
and Technology of Japan and by a Grand-in-Aid for the 21st Century 
COE ``Center for Diversity and Universality in Physics".

\end{document}